**Cultural Differences in Students' Privacy Concerns in Learning Analytics across Germany, South Korea, Spain, Sweden, and the United States**

Olga Viberg[1*], René F. Kizilcec[2*], Ioana Jivet[3], Alejandra Martínez Monés[4], Alice Oh[5], Chantal Mutimukwe[6], Stefan Hrastinski[7], Maren Scheffel[8]

[1]School of Electrical Engineering and Computer Science, KTH Royal Institute of Technology, Lindstedsvägen 3, 10044 Stockholm, Sweden. Email: oviberg@kth.se

[2]Department of Information Science, Cornell University, 107 Hoy Rd, Ithaca, NY 14853, USA. Email: kizilcec@cornell.edu

[3]DIPF Leibniz Institute for Research and Information in Education, Rostocker Straße 6, 60323 Frankfurt am Main, Germany. Email: jivet@dipf.se

[4]Scuela de Ingeniería Informática, Valladolid University, Paseo de Belén, 15, 47011 Valladolid, Spain. Email: amartine@infor.uva.es

[5]School of Computing, The Korea Advanced Institute of Science and Technology (KAIST), 291 Daehak-ro, Yuseong-gu, Daejeon, Korea 34141. Email: alice.oh@kaist.edu

[6]Department of Computer and Systems Science, Stockholm University, NOD-huset, Borgarfjordsgatan 12, 164 55 Kist, Stockholm, Sweden. Email: chantal.mutimukwe@dsv.su.se

[7]Department of Learning and Engineering Sciences, KTH Royal Institute of Technology, Osquars backe 31, 10044 Stockholm, Sweden. Email: stefanhr@kth.se

[8]Institut für Erziehungswissenschaft, Ruhr-Universität Bochum, Universitätsstraße 150, D - 44801 Bochum, Germany. Email: maren.scheffel@rub.de

[*]Authors contributed equally.

Correspondence to Olga Viberg (oviberg@kth.se) and Rene Kizilcec (kizilcec@cornell.edu).

**Abstract**

Applications of learning analytics (LA) can raise concerns from students about their privacy in higher education contexts. Developing effective privacy-enhancing practices requires a systematic understanding of students' privacy concerns and how they vary across national and cultural dimensions. We conducted a survey study with established instruments to measure privacy concerns and cultural values for university students in five countries (Germany, South Korea, Spain, Sweden, and the United States; $N = 762$). The results show that students generally trusted institutions with their data and disclosed information as they perceived the risks to be manageable even though they felt somewhat limited in their ability to control their privacy. Across the five countries, German and Swedish students stood out as the most trusting and least concerned, especially compared to US students who reported greater perceived risk and less control. Students in South Korea and Spain responded similarly on all five privacy dimensions (perceived privacy risk, perceived privacy control, privacy concerns, trusting beliefs, and non-self-disclosure behavior), despite their significant cultural differences. Culture measured at the individual level affected the antecedents and outcomes of privacy concerns. Perceived privacy risk and privacy control increase with power distance. Trusting beliefs increase with a desire for uncertainty avoidance and lower masculinity. Non-self-disclosure behaviors rise with power distance and masculinity, and decrease with more uncertainty avoidance. Thus, cultural values related to trust in institutions, social equality and risk-taking should be considered when developing privacy-enhancing practices and policies in higher education.

## 1. Introduction

Learning analytics have the potential to improve learning and teaching, as well as to improve decision making in education (Ifenthaler et al., 2021). While learning analytics technology has matured in the last decade, large-scale adoptions are still limited in practice (Gašević et al., 2019; Tsai et al., 2018; Viberg et al., 2018). The opportunities that learning analytics can provide for improving learning and teaching are often linked to critical privacy issues that may impede the successful implementation of learning analytics services at scale (Li et al., 2022). In fact, data privacy has been found to be a major concern for learning analytics research and practice, and for advancing educational research more generally (Ferguson, 2012; Hoel & Chen, 2018; Joksimovic et al., 2022; Kimmons, 2021; Li et al., 2022; Potgieter, 2021; Selwyn, 2019). Whereas the importance of effectively addressing students' privacy concerns has been continuously raised by the learning analytics community over the last decade (e.g., Joksimovic, 2022, Pardo & Siemens, 2014; Prinsloo & Slade, 2015; Prinsloo et al., 2022), privacy issues, including concerns are rarely addressed in learning analytics practice (Mutimukwe et al., 2022; Priedigkeit et al., 2021). This may be due to our limited understanding of these concerns, which refer to individual worries about the possible loss of privacy resulting from information disclosure to a specific external agent or institution (Xu et al., 2011). Mutimukwe and colleagues have recently proposed the SPICE model (Students' Information Privacy Concerns) to uncover antecedents (perceived privacy risk and perceived privacy control) and outcomes (trusting beliefs and non-self-disclosure behavior) of students' privacy concerns about learning analytics (Mutimukwe et al., 2022). However, the SPICE model does not consider contextual factors, including national and cultural influences of privacy concerns.

Building on the general SPICE model, we examine these additional factors that may play an important role in students' privacy concerns linked to the use of learning analytics services. As stated by Milberg et al. (1995), regulations and policies regarding the use of personal information differ from one country to another, "as may the nature and level of information privacy concern" (p.66). Therefore, "understanding the differences in information privacy concerns [...] may be a key to successfully managing these concerns" (p.66). Research focusing on the understanding of individuals' information privacy concerns in other disciplines has shown associations between a number of individuals' cultural values (e.g., power distance and individualism) and information privacy issues (Milberg et al., 1995) across countries. For example, countries that score highly on the *individualism* dimension (i.e., where people emphasize individual initiative and achievement) tend to put more value on individuals' private life as opposed to the countries with low scores in individualism, in which "there is more of an acceptance that organizations will invade one's private life" (Milberg et al., 1995, p.68). Other studies showed that countries with a high *power distance* index (i.e., where there is a high degree of inequality between a less powerful entity and a more powerful one) demonstrate lower levels of trust (Hofstede, 1980). Further, scholars found that people from cultures ranking high in *uncertainty avoidance* (i.e., the extent to which the members of a culture feel threatened by ambiguous and unknown situations) found privacy risks to be more important when making privacy-related disclosure decisions (Trepte et al., 2017). In the context of learning analytics, Hoel and Chen (2018) highlight that "there are clear differences in the way privacy is conceptualised" (p.4) across countries. As an example, they stress that concerns about the rights of the individual in relation to control of data emanating from the learner represent a Western tradition, as compared to the East, where the individual's interests are more frequently projected onto the group's interest.

Considering the importance of the understanding of cross-cultural differences in students' information privacy concerns in the setting of learning analytics, we ought to better understand how their privacy concerns vary across countries and cultures. Specifically, we address the following research questions: *How do students' privacy concerns in learning analytics vary across countries* (**RQ1**), and *How does culture affect these privacy concerns* (**RQ2**)? This empirical study contributes relevant evidence by examining students' information privacy concerns in learning analytics across five countries: Germany, South Korea, Spain, Sweden, and the United States in a higher education setting. We selected countries in five distinct geographical regions (North, Central, and South Europe; Southeast Asia, and North America), in which people's cultural values, as measured earlier by Hofstede et al. (2010) have been found to differ along several dimensions (e.g., power distance varies from 31 in

Sweden to 60 in South Korea, and individualism varies from 18 in South Korea to 91 in the USA; see Table 1).

## 2. Background

### 2.1. Privacy concerns in learning analytics

Privacy is an essential part of ethical learning analytics practice (Marshall et al., 2022) and a topic of interest among researchers in several disciplines (Smith et al., 2011), who have defined privacy in different ways. While it is often defined as a legally established right to be left alone (Warren & Brandies, 1890), or limited access or isolation in philosophy and psychology (Schoeman, 1984), researchers in information systems and social sciences have suggested that privacy is one's ability to control information about oneself (Margulis, 2003; Westin, 1967). Privacy has been described as multidimensional, elastic, dependent upon context (Bélanger & Crossler, 2011), situation (Xu et al., 2011), and cultural values (Bellman et al., 2004). Due to the complexity of quantifying privacy, social scientists have relied on measuring privacy-related proxies, and there has been a push for focusing on 'privacy concerns' as the central construct (Xu et al., 2011).

Privacy in learning analytics has been studied through different measures such as stakeholders' expectations of privacy-related issues (Viberg et al., 2022; Whitelock-Wainwright et al., 2020) and stakeholders' perceptions (Ifenthaler & Schumacher, 2016), awareness (Velander, 2020), preferences (Jones, 2019; Korir et al., 2022), perspectives (Jones et al., 2020), and attitudes towards privacy (Slade et al., 2019). For example, the Data Doubles project at Indiana University Indianapolis is a three-year effort to investigate students' perspectives on privacy issues associated with academic library participation in LA initiatives through interviews, surveys, and focus groups (Jones et al., 2023). Privacy concerns in learning analytics clearly warrant in-depth investigation. Privacy concerns refers to individuals' concerns about the possible loss of privacy that would result from disclosing information to a specific agent or institution (Xu et al., 2011). This differentiates it from constructs that focus on individuals' expectations, perceptions, or awareness of how institutions should handle their personal information (Hong & Tong, 2013). For instance, a student may expect the university to protect her personal data, but she may still have related privacy concerns since she may not fully trust the university in meeting the expectations.

Earlier research on technology usage and adoption of information systems showed that information privacy concerns affect the adoption of various technologies due to the desire not to disclose personal information (Lowry et al., 2011). Thus, understanding students' privacy concerns is a critical step towards developing effective privacy-enhancing practices in learning analytics (Ahn et al., 2021; Mutimukwe et al., 2021; Tsai et al., 2020) since this may influence the successful adoption of learning analytics systems. In this work, we build on a model of student privacy concerns developed by Mutimukwe et al. (2022) based on a review of the literature (the SPICE model is reviewed in Section 3). Other studies have addressed this issue indirectly. For instance, the literature review of learning analytics dashboards by Williamson and Kizilcec (2022) identified six papers where learners' privacy concerns had emerged even though these concerns were not the focus of these studies: for example, students raised concerns about their privacy being exposed in a dashboard study (Roberts et al., 2017). Other researchers identified potential threats from learning analytics strengthening the power imbalance between learners and instructors due to increased monitoring capabilities (Han et al., 2021; Wise & Jung, 2019), and highlighted the importance of better understanding cross-cultural differences in students' information privacy concerns about learning analytics (Hoel & Chen, 2019). However, to the best of our knowledge, few studies have examined privacy concerns in learning analytics across countries. This motivates our first research question:

**RQ1:** How do students' privacy concerns in learning analytics vary across countries?

### 2.2. Culture and learning analytics

Learning analytics applications have been designed, implemented and used across countries in many ways (e.g., Viberg et al., 2018). Furthermore, teachers and students have different expectations towards learning analytics in different countries (e.g., Kollom et al., 2021; Hilliger et al. 2020; Pontual Falcao et al., 2022; Viberg et al., 2022) and different concerns about the ethical and privacy-related issues around learning analytics (Hoel & Chen 2019; West et al., 2020). These contextual, technical, and also cultural differences make the transfer of learning analytics applications across countries challenging. Whereas technical and contextual aspects of learning analytics design and implementation have been addressed by the learning analytics community, cultural factors have hitherto received little attention (Jivet et al., 2022), even though the importance of addressing culture in learning analytics has been raised by researchers over a decade ago (Vatrapu, 2011). In particular, Vatrapu (2011) suggested that learning analytics should consider culture in both appropriation of affordances (tool use) and technological intersubjectivity (how students and teachers relate to, interact with, and form impressions of each other in technology-enhanced learning settings). Researchers have studied cultural differences in the use and impact of learning analytics applications (e.g., Cho et al., 2021; Davis et al., 2017; Kizilcec & Cohen, 2017; Mittelmeier et al., 2016), but not the influence of culture on students' privacy concerns.

## 2.3. Cultural differences in information privacy concerns

Prior research found that cultural values differ across countries, which can affect a society's response to the environment, and "may also be associated with individual privacy concerns" (Milberg et al., 2000, p.39). Milberg et al. (2000) combined four of Hofstede's (1980, 1991) cultural indices, *power distance*, *individualism*, *masculinity*, and *uncertainty avoidance*, into an overall measure of cultural values, and found that it significantly predicted information privacy concerns in information systems across countries. Concerns about information privacy were positively associated with *power distance*, individualism and *masculinity*, and negatively associated with *uncertainty avoidance*. Milberg et al. (2000) offered an explanation for these correlations: although cultures with a high *power distance* index tolerate greater levels of inequality in power, higher scores are associated with greater mistrust of more powerful groups, such as organizations or institutions. Low *individualism* (i.e., collectivist) societies have a greater acceptance that groups, including organizations, can intrude on the private life of the individual. High *masculinity* cultures place greater emphasis on achievement and material success, and perhaps the economic benefits of using private information, over caring relationships and quality of life. Finally, societies with a high *uncertainty avoidance* index tend to reduce uncertainty by embracing clear written rules and regulations, and may be more likely to introduce higher levels of government regulation of privacy. Other research has also shown links between several individual values, including cultural ones and information privacy concerns (Bellman et al., 2004; Smith et al., 1996; Stone & Stone, 1990). This motivates our second research question:

> **RQ2:** How does culture impact students' privacy concerns in learning analytics?

## 3. Theoretical lens

The theoretical model that we use in this study to examine students' privacy concerns in learning analytics is the SPICE model (Mutimukwe et al., 2022). To examine culture, we use Hofstede's framework of culture (Hofstede et al., 2010). In this section, we briefly present these two established theoretical frameworks that we build on in this study.

## 3.1. The SPICE model

The students' privacy concerns (SPICE) model has been developed and empirically validated to explore the nature of students' privacy concerns in learning analytics in higher education (Mutimukwe et al., 2022). The SPICE model considers students' privacy concerns as a central construct between two antecedents – perceived privacy risk and perceived privacy control, and two outcomes – trusting beliefs and non-self-disclosure behavior (Figure 1). Overall, it explores how students' risk-control perceptions may influence their privacy concerns, trusting beliefs as well as non-self-disclosure behavior. Correspondingly, the model describes the directional relationships between students' privacy concerns, trusting beliefs, and non-self-disclosure behavior.

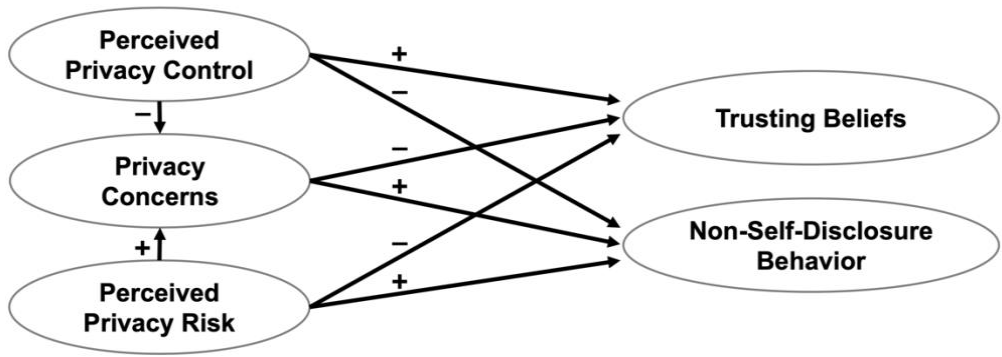

Fig. 1: Model of Students' Privacy Concerns (SPICE) indicating theorized relationships between constructs (Mutimukwe et al., 2022).

The SPICE model is grounded in the Antecedents-Privacy Concerns-Outcomes (APCO) framework proposed by Smith et al. (2011), which provides a comprehensive view of the link between privacy concerns, its antecedents, and its outcomes. *Privacy concerns* are explained as individual concerns about the possible loss of privacy resulting from information disclosure to a specific external agent or institution (Xu et al., 2011). *Perceived privacy risk* refers to "the perceived potential risk when personal information is revealed" (Dinev & Hart, 2004, p. 415). It has been considered as a factor that influences the perceived state of privacy and individual experience (e.g., Petronio, 2002). *Perceived privacy control* refers to the individual's beliefs in her/his ability to manage the release and dissemination of personal information (Malhotra et al., 2004). *Trusting* beliefs are considered at the degree to which higher education institutions are dependable in protecting students' personal information (Malhotra et al., 2004). *Non-self-disclosure behaviors* involve revealing information about oneself to others (Derlega et al., 1993).

### 3.2. Hofstede's model of national culture and privacy concerns

Hofstede defines culture as "the collective programming of the mind which distinguishes the members of one group or category of people from another" (Hofstede et al., 2010, p.5). Culture is often studied and understood through cultural values: a set of strongly held beliefs that guide attitudes and behavior of a group or society and tend to endure even when other differences between countries are eroded by changes in economics, politics, technology, and other external pressures (Engel et al., 1986; Hofstede, 1980; Long & Quek, 2002). We examine students' cultural values using Hofstede's dimensions of national cultures around the world (Hofstede et al., 2010), which have been argued to "explain human behaviors better than other measures, such as country and language" (Li, 2022, p. 269). Hofstede's model, based on data from over 100,000 IBM employees in 40 countries, has been used across research fields and studies, including in the context of education (e.g., Huang et al., 2019; Tarhini et al., 2017) and information privacy concerns across cultures (e.g., Li, 2022; Li et al., 2022; Milberg et al., 2000).

Hofstede's model has been used extensively and across multiple domains, but not without significant critique (e.g., Baskerville, 2003; Bhimani et al., 2005; Harrison & McKinnon, 1999; McSweeney, 2002). This critique comprises theoretical and empirical weaknesses (Joannidés et al., 2012): first, equating cultures with nations is incorrect; second, conceptualizing cultures as fixed, rather than changing and flexible, is incorrect; and third, the underlying data were collected in IBM offices in the 1960-70s, not in internationally mixed today's higher education contexts (Signorini et al., 2009). However, despite these valid criticisms, Hofstede's model has been found to be useful in exploring, for example, individual's technology acceptance (e.g., Huang et al., 2019; Teo & Huang, 2018), trust in recommendation systems (e.g., Berkovsky et a., 2018), and individual information privacy concerns in information systems (Jones & Alony, 2007; Li, 2022; Milberg et al., 2000). To address concerns about equating cultures to nations, we also use Hofstede's dimensions (see Section 3.2) to unpack students' cultural values at the *individual level*, following Yoo et al. (2011).

While we use Hofstede's model of national culture (Hofstede et al., 2010) as a proxy for culture in this study, we do not suggest that cultural dimensions are directly representing nationalities or that nations are homogenous in terms of cultural characteristics and values. Instead, we use the four dimensions that were found to be critical for individuals' information privacy concerns in information systems (Milberg et al., 1995, 2000) to represent certain cultural values that we expect to affect students' privacy concerns in learning analytics: *individualism/collectivism*, *masculinity/femininity*, *power distance*, and *uncertainty avoidance*. In describing how these dimensions might manifest in an educational setting, Hofstede et al. (2010) depict very similar scenarios to the explanations provided by Milberg (2000) for the associations between privacy concerns and these four dimensions. For example, in the low *power distance* societies "students treat teachers as equals" and "expect initiatives from students in class", and in high *power distance* societies, the distance between teachers and students is high, even outside the classroom and teachers are expected to take all initiatives in class (Hofstede et al., 2010, p.72). Table 1 presents the definition and original index score for each of the four dimensions for Germany, South Korea, Spain, Sweden, and the United States.

**Table 1.** Hofstede's cultural dimensions and the index scored for Germany, South Korea, Spain, Sweden, and the USA (Hofstede et al., 2010).

| Dimensions | Germany | South Korea | Spain | Sweden | USA |
|---|---|---|---|---|---|
| *Power distance* | *Power distance* describes "the extent to which the less powerful members of institutions and organisations within a country expect and accept that power is distributed unequally" (p.61). | | | | |
| | 35 | 60 (large power distance) | 57 | 31 | 40 |
| *Individualism / Collectivism* | *Individualism* describes "societies in which the ties between individuals are loose: everyone is expected to look after themselves and their immediate family" (p. 92). *Collectivism* describes "societies in which people from birth onward are integrated into strong, cohesive in-groups, which throughout people's lifetime continue to protect them in exchange for unquestioning loyalty" (p. 92). | | | | |
| | 67 | 18 | 51 | 71 | 91 (very individualistic) |
| *Masculinity/ Femininity* | A *masculine* society is one where "emotional gender roles are clearly distinct: men are supposed to be assertive, tough, and focused on material success, whereas women are supposed to be more modest, tender, and concerned with the quality of life" (p. 140). In a *feminine* society, "emotional gender roles overlap: both men and women are supposed to be modest, tender, and concerned with the quality of life." (p. 140) | | | | |
| | 66 (very masculine) | 39 | 42 | 5 | 62 |
| *Uncertainty avoidance* | *Uncertainty avoidance* describes "the extent to which the members of a culture feel threatened by ambiguous and unknown situations" (p. 191). | | | | |
| | 65 | 85 | 86 (strong desire for uncertainty avoidance) | 29 | 46 |

## 4. Method

### 4.1. Participants and context

All five samples were collected from university students, but the sample collection process varied across universities to adhere to the local standards. Institutional ethical approval was obtained in advance of data collection, and all respondents included in the analysis provided informed consent. Students' participation was voluntary at all locations. Data collection occurred between November 2021 and August 2022. The US sample was collected at a selective research university on the East Coast, using a cloud-based participant management system that offers course credit in return for participation in research studies. The system is used primarily by students pursuing communication or information science degrees. Students completed the survey online using Qualtrics. The South Korean sample was collected at a selective, STEM-focused university in South Korea. Instructors of several large lecture courses announced the survey study for voluntary participation to their students by email. Students completed the survey online using Qualtrics. The Spanish sample was collected at a medium-size, traditional university in the Spanish inner plateau. Data was collected via a Microsoft Forms online survey that was announced to students as an institutional message. The Swedish sample was collected at a large technical university in Sweden, focused on engineering disciplines. Data was collected via a survey distributed to students in paper format. The survey distribution adhered to recommendations from the institutional ethical pre-approval board and followed guidelines from the Swedish National Ethical Board[1]. The German sample was collected using LimeSurvey and participants were recruited through prolific.co[2], an online platform for recruiting subjects for online studies (Palan & Schitter, 2018). The invitation to participate in the study was distributed to students with German nationality currently living in Germany.

The sample characteristics for each country and the full sample are provided in Table 2. Most participants were between 18 and 24 years old, enrolled in a Bachelor's or Master's program in a STEM or HASS field, with a skew towards female students except in South Korea and Sweden. We examined respondents' self-reported nationality and country of residence, which in the vast majority of cases corresponded to the country in which the sample was collected; only the US sample included a notable share of respondents whose nationality was East Asian, which reflects the university's overall enrolment statistics.

**Table 2.** Participant characteristics in each sample and overall.

|  | **Germany** | **South Korea** | **Spain** | **Sweden** | **USA** | **Overall** |
|---|---|---|---|---|---|---|
| **Sample Size** | 247 (32%) | 59 (8%) | 121 (16%) | 156 (20%) | 179 (23%) | 762 (100%) |
| **Sex** <br> Female <br> Male <br> Other/NA | <br> 142 (57%) <br> 101 (41%) <br> 4 (2%) | <br> 17 (29%) <br> 41 (69%) <br> 1 (2%) | <br> 84 (69%) <br> 36 (30%) <br> 1 (1%) | <br> 75 (48%) <br> 76 (49%) <br> 5 (3%) | <br> 117 (65%) <br> 59 (33%) <br> 3 (2%) | <br> 435 (57%) <br> 313 (41%) <br> 12 (2%) |
| **Age** <br> 18-24 <br> 25-34 <br> 35+ | <br> 142 (57%) <br> 91 (37%) <br> 14 (6%) | <br> 43 (73%) <br> 14 (24 %) <br> 2 (3%) | <br> 90 (74%) <br> 20 (17%) <br> 11 (9%) | <br> 126 (81%) <br> 29 (19%) <br> 1 (0%) | <br> 175 (98%) <br> 3 (2%) <br> 1 (0%) | <br> 576 (76%) <br> 157 (21%) <br> 29 (4%) |
| **Program** <br> BA <br> MA <br> Indiv. Course/other | <br> 145 (57%) <br> 77 (31%) <br> 25 (10%) | <br> 39 (66%) <br> 18 (31%) <br> 2 (3%) | <br> 105 (87%) <br> 16 (13%) <br> 0 (0%) | <br> 56 (36%) <br> 84 (54%) <br> 16 (10%) | <br> 168 (94%) <br> 6 (3%) <br> 5 (3%) | <br> 513 (67%) <br> 201 (26%) <br> 48 (6%) |

---

[1] https://etikprovningsmyndigheten.se  
[2] https://www.prolific.co/

| **Field of study** | | | | | | |
|---|---:|---:|---:|---:|---:|---:|
| STEM* | 79 (32%) | 56 (95%) | 46 (38%) | 145 (93%) | 88 (49%) | 414 (54%) |
| HASS§ | 144 (58%) | 2 (3%) | 65 (54%) | 6 (4%) | 62 (35%) | 279 (37%) |
| Medical | 12 (5%) | 1 (2%) | 10 (8%) | 0 (0%) | 1 (1%) | 24 (3%) |
| Other | 12 (5%) | 0 (0%) | 0 (0%) | 5 (3%) | 28 (16%) | 45 (6%) |

*Notes:* *STEM refers to Science, Technology, Engineering, and Mathematics. §HASS refers to Humanities, Arts, and Social Sciences.

### 4.2. Measures

Data was collected using a survey instrument that consisted of three parts (Appendix). The first part contained questions about student demographic information, such as their age, gender, year of study, subject area, and the type of degree program (e.g., bachelor or master levels). The second part contained twenty items about the nature of students' privacy concerns in the context of learning analytics, adopted from the validated instrument, developed by Mutimukwe et al. (2022). The instrument consists of the five subscales with four items each that are reviewed in Section 3: perception of privacy control (e.g., "I believe I have control over how my personal information is used by my university."), perceived risks (e.g., "In general, it would be risky to give personal information to the learning management system used by my university."), privacy concerns (e.g., "I am concerned that the information I provide to my university could be misused."), trusting beliefs (e.g., "I trust that my university tells the truth and fulfils promises related to my personal information."), and non-self-disclosure behavior (e.g., "I refuse to use the learning management system because I disagree with the way my university uses personal information.").

The final part of the survey instrument contained twenty questions about cultural values based on four of Hofstede's cultural categories (power distance, individualism vs. collectivism, uncertainty avoidance and masculinity vs. femininity). The items were adopted from Yoo et al.'s (2011) *CVSSCALE (Individual Cultural Values Scale)*, which measures individual cultural values at the *individual level*, in contrast to the cultural values measured at a national level by Hofstede (2010). The measure we used was found to be valid in both "student and nonstudents samples, which also indicates cross-sample generalizability" in five countries (Yoo et al., 2011, p. 205). Participants rated the items on 5-point Likert scales.

The original survey instrument was translated from English into German, South Korean, Spanish, and Swedish. Translation into South Korean was conducted by a Korean native speaking research assistant, and subsequently confirmed by a second native speaker, one of the authors of this study. Translation into Spanish was initially conducted by a Spanish PhD student, and later checked by one of the authors, who is a native speaker of Spanish. Translation into Swedish was conducted by two lecturers of Swedish as a second language teaching Swedish in a higher educational setting. The translated version was piloted with two Swedish students before it was distributed to a larger sample. Small revisions to three statements were made based on these pilot results. Translation into German was conducted by a German native speaker and double-checked by a second native speaker. The result was translated back to English by a bilingual, native English speaker and adjustments were made to four items.

### 4.3. Analysis

Survey responses from the five countries were combined into a single file, excluding responses from individuals who either did not provide informed consent for study participation (n=6) or who responded to less than half of the relevant questions (n=16 responses, all from South Korea). This final sample included 247 responses from Germany, 59 from South Korea, 121 from Spain, 156 from Sweden, and 179 from the US. A small percentage of missing values for items that appeared later in the survey (<3%) were imputed using multiple imputation through predictive mean matching with the mice *R* package (Van Buuren & Groothuis-Oudshoorn, 2011).

Following the imputation step, we aggregated the subscales for our key constructs and checked the internal reliability of the subscales overall and in specific cultural contexts (Table 3). All responses were collected on five-point Likert scales and coded from 1 (strongly disagree) to 5 (strongly agree). The subscales for our key constructs show sufficient internal reliability, variance, and no floor or ceiling effects; moreover, the level of internal reliability is highly consistent across individual samples for most constructs. Notable exceptions are lower internal reliability for *power distance* in Sweden (alpha=0.53) and *masculinity* in South Korea (alpha=0.023). The low internal reliability of the *masculinity* subscale in the Korean sample is largely due to one item ("There are some jobs that a man can always do better than a woman.") that respondents tend to agree with despite generally disagreeing with the other items in the subscale. We confirmed the accuracy of the translation and confirmed with a native Korean that this response pattern is reasonable in light of recent debate about the suitability of women for professions such as in law enforcement (Jun, 2021).

**Table 3.** Mean, standard deviation (in parentheses), and internal reliability (Cronbach's alpha) for each construct in each sample and overall.

|  | Germany | South Korea | Spain | Sweden | USA | Overall |
|---:|---|---|---|---|---|---|
| Perc. Privacy Control | 3.0 (0.80), a=0.85 | 2.8 (1.0), a=0.88 | 2.8 (0.83), a=0.83 | 2.7 (0.85), a=0.82 | 2.6 (0.81), a=0.84 | 2.8 (0.85), a=0.85 |
| Perc. Privacy Risk | 2.5 (0.73), a=0.73 | 2.6 (0.89), a=0.82 | 2.7 (0.86), a=0.86 | 2.3 (0.72), a=0.74 | 3.0 (0.72), a=0.70 | 2.6 (0.80), a=0.78 |
| Privacy Concerns | 2.3 (0.85), a=0.86 | 2.8 (1.0), a=0.89 | 2.9 (0.97), a=0.89 | 2.0 (0.80) a=0.84 | 3.1 (0.85), a=0.83 | 2.6 (0.96), a=0.88 |
| Trusting Beliefs | 4.1 (0.60), a=0.84 | 3.4 (0.86), a=0.88 | 3.6 (0.86), a=0.88 | 4.0 (0.80), a=0.88 | 3.3 (0.81), a=0.87 | 3.8 (0.83), a=0.88 |
| Non-Self-Disclosure Behavior | 1.6 (0.58), a=0.77 | 2.0 (0.79), a=0.80 | 2.0 (0.81), a=0.87 | 1.3 (0.46), a=0.81 | 2.1 (0.8), a=0.86 | 1.7 (0.75), a=0.85 |
| Power Distance | 1.6 (0.55), a=0.75 | 1.6 (0.66), a=0.81 | 1.7 (0.72), a=0.83 | 1.7 (0.49), a=0.53 | 1.8 (0.70), a=0.86 | 1.7 (0.62), a=0.78 |
| Individualism | 2.9 (0.66), a=0.80 | 2.1 (0.61), a=0.73 | 3.2 (0.78), a=0.85 | 2.8 (0.69), a=0.80 | 3.0 (0.62), a=0.77 | 2.9 (0.72), a=0.82 |
| Masculinity | 1.8 (0.88), a=0.83 | 2.3 (0.49), a=0.023 | 1.7 (0.74), a=0.76 | 1.5 (0.64), a=0.67 | 1.9 (0.79), a=0.76 | 1.8 (0.79), a=0.75 |
| Uncertainty Avoidance | 3.7 (0.49), a=0.72 | 4.1 (0.65), a=0.86 | 3.7 (0.53), a=0.65 | 3.7 (0.59), a=0.74 | 3.9 (0.58), a=0.83 | 3.8 (0.56), a=0.76 |

## 5. Results

The average subscale scores presented in Table 3 show that the average student in our sample had limited concerns about their privacy in learning analytics and the perceived risks, and they maintained trusting beliefs and an inclination towards self-disclosure, despite perceiving their privacy control as somewhat limited. The overall average score was at or below the neutral midpoint (3) for perceived privacy control (i.e., less perceived control), privacy concerns (i.e., less concerned), perceived privacy risk (i.e., lower perceived risk), and non-self-disclosure behavior (i.e., more inclined to self-disclose), and above the midpoint for trusting beliefs (i.e., more trusting).

**5.1. Students' privacy concerns in learning analytics vary across countries**

We find differences in privacy concerns between students in different countries based on the means and standard errors for each construct, as shown in Figure 2. We conducted a set of non-parametric rank sum tests that confirm significant sample differences between countries for each construct (all Kruskal-Wallis $X^2_{df=4} > 34.0$, p-values $< 0.001$). Next, we examine which country samples differ for each construct using the same non-parametric Kruskal-Wallis test (t-tests yield qualitatively equivalent results) and percentage differences in group means. Perceived privacy control is 12% higher in the German sample relative to the samples from the other four countries ($X^2_{df=1} = 26.8$, $p < 0.001$), but it does not significantly differ between the Korean, Spanish, Swedish, and the US samples ($X^2_{df=3} = 7.10$, $p = 0.069$). Perceived privacy risk is 17% higher in the US sample ($X^2_{df=1} = 38.5$, $p < 0.001$) and 12% lower in the Swedish sample compared to the German, Korean, and Spanish samples ($X^2_{df=1} = 13.7$, $p < 0.001$), where it does not differ significantly ($X^2_{df=2} = 3.50$, $p = 0.174$). Privacy concerns are lowest in the Swedish sample, 13% lower than the second lowest from Germany ($X^2_{df=1} = 14.7$, $p < 0.001$); privacy concerns in the German sample are 11% lower than in the Korean, Spanish, and the US samples ($X^2_{df=1} = 67.4$, $p < 0.001$), where they do not differ significantly ($X^2_{df=2} = 5.70$, $p = 0.058$). Trust beliefs are similarly high in the German and Swedish samples ($X^2_{df=1} = 1.57$, $p = 0.209$), and 7% higher than in the Spanish sample ($X^2_{df=1} = 25.0$, $p < 0.001$). Trusting beliefs are similarly low in the US and Korean samples ($X^2_{df=1} = 1.35$, $p = 0.245$) but 10% higher in the Spanish sample compared to the US sample ($X^2_{df=1} = 37.1$, $p < 0.001$). Finally, non-self-disclosure behaviors are 22% lower in the Swedish relative to the German sample ($X^2_{df=1} = 55.9$, $p < 0.001$), and 22% lower in the German than in the Korean, Spanish, and the US samples ($X^2_{df=1} = 49.4$, $p < 0.001$), where they do not differ significantly ($X^2_{df=2} = 4.05$, $p = 0.132$).

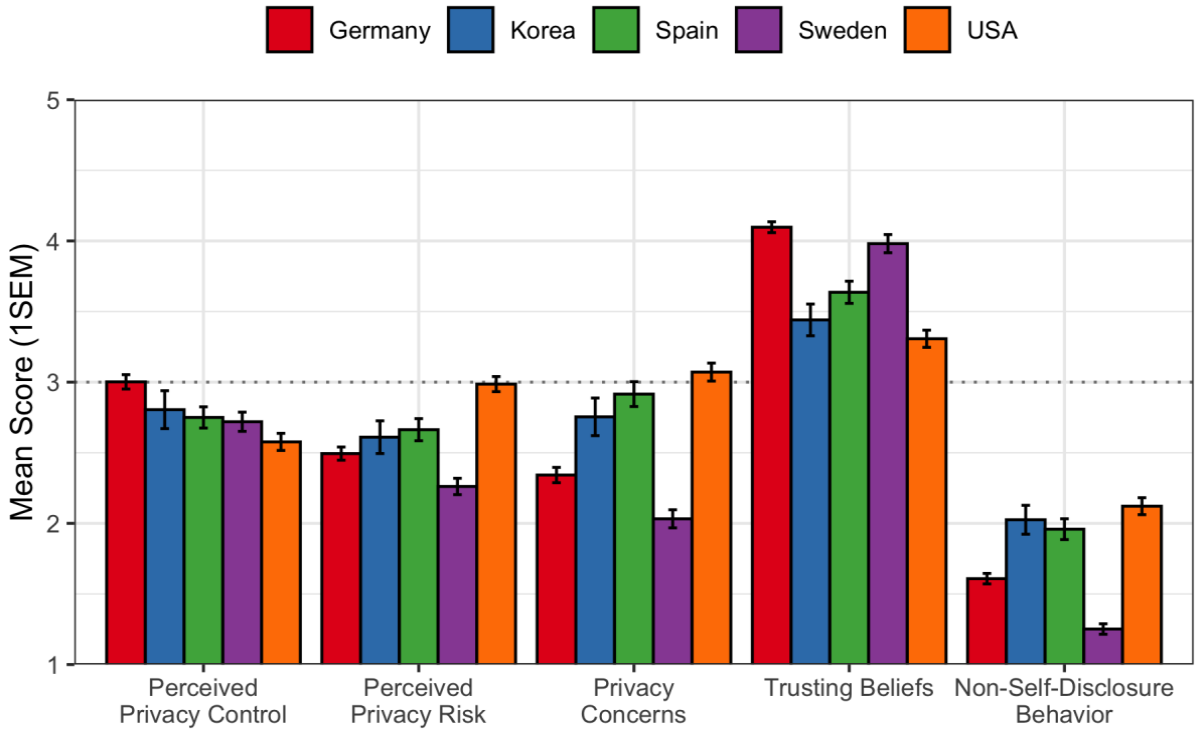

Fig. 2. Mean and standard error for each construct of student privacy concerns for each country.

**5.2. Culture affects students' privacy concerns in learning analytics**

To study how culture affects students' privacy concerns, we examine how culture affects students' privacy concerns using *individual-level* responses for Hofstede's four cultural dimensions. This enables us to examine relationships across the five samples to identify the influence of each cultural dimension. Figure 3 shows the average response for each cultural dimension in each sample. These deviate from the Hofstede country-level scores in Table 1 because the underlying samples are drawn from different populations and at different times. For example, the Spanish sample is the most individualistic at the individual level but the USA is the most *individualistic* country at the country level; and while the South Korean and Spanish samples differ in *masculinity* and *uncertainty avoidance* measured at the individual level, they are almost identical at the country level. These observations underline the need for an individual-level analysis of cultural differences to accurately capture the cultural values of students in our samples.

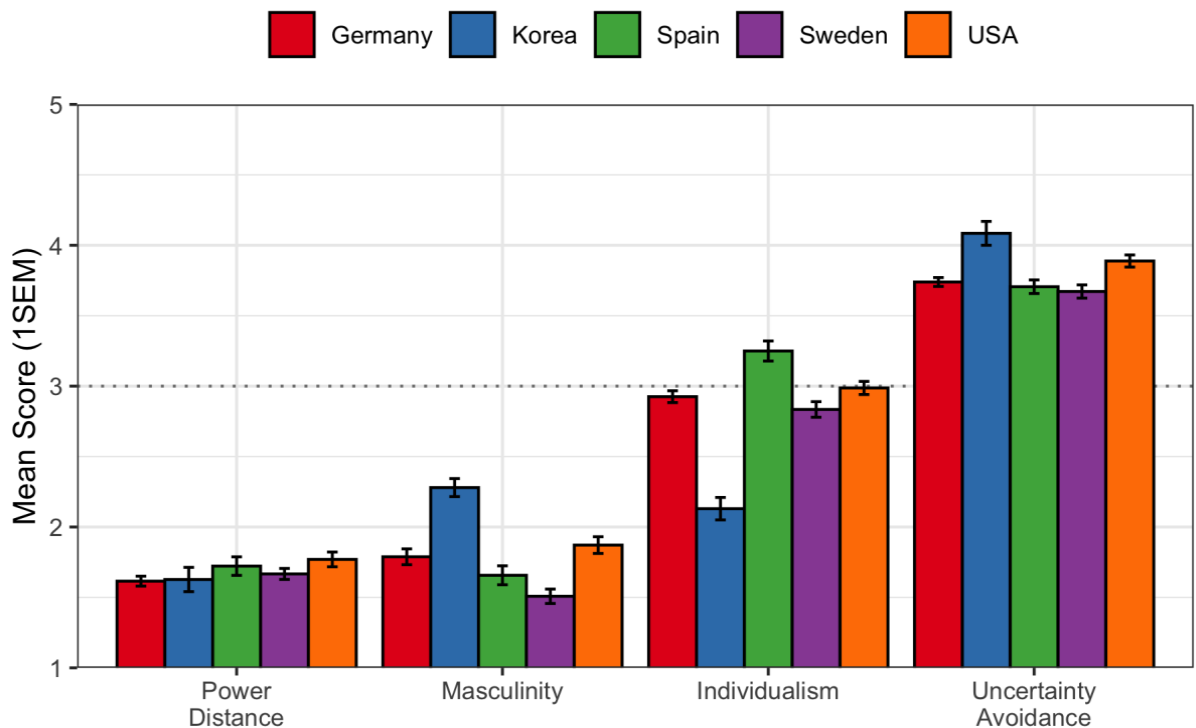

Fig. 3. Means with standard error bars for each individually measured cultural dimension in each sample.

To examine how culture influences students' privacy concerns in LA at the individual level, we regress responses for each privacy concern construct on responses for all four cultural dimensions, adding sample fixed effects to control for aggregate country-level differences. We confirmed that there is no multicollinearity among the independent variables (VIF < 1.4), and we use heteroskedasticity-robust standard errors. Results of multiple regression analysis are presented in Table 4. The significant F-statistics for each regression model indicate that cultural values and the fixed effects overall explain a significant amount of variance in privacy concerns in LA: cultural differences explain different amounts of variance across the five constructs, ranging from 31% for non-self-disclosure behavior to 5% for perceived privacy control. Our focus in this analysis is however on the estimated coefficients that indicate the strength of the relationship between each cultural value and privacy concern. We find that *power distance* is a significant positive predictor of all privacy concern constructs except trusting beliefs and marginally for privacy concerns. *Masculinity* is a significant positive predictor of non-self-disclosure behavior but a negative predictor of trusting beliefs, whereas *uncertainty avoidance* is a significant negative predictor of non-self-disclosure behavior and a positive predictor of trusting beliefs. These regression coefficients are nearly identical when excluding the Korean sample, where the

*masculinity* score was not internally consistent. *Individualism* is not a significant predictor for any privacy construct.

**Table 4.** Multiple regression analysis for each student privacy concern regressed on four individual-level cultural dimensions with sample fixed effect. Unadjusted regression coefficients are shown with robust standard errors in parentheses.

|  | Perceived Privacy Control | Perceived Privacy Risk | Privacy Concerns | Trusting Beliefs | Non-Self-Disclosure Behavior |
|---|---|---|---|---|---|
| Power Distance | 0.12* (0.06) | 0.12* (0.06) | 0.13 (0.07) | 0.01 (0.06) | 0.30*** (0.05) |
| Masculinity | 0.06 (0.05) | 0.08 (0.04) | 0.07 (0.05) | -0.10* (0.05) | 0.14*** (0.04) |
| Individualism | 0.09 (0.05) | -0.03 (0.05) | -0.02 (0.06) | 0.08 (0.05) | -0.02 (0.04) |
| Uncertainty Avoidance | 0.01 (0.06) | -0.03 (0.05) | 0.04 (0.06) | 0.23*** (0.06) | -0.17*** (0.04) |
| (Intercept) | 2.47*** (0.27) | 2.38*** (0.25) | 1.93*** (0.28) | 3.18*** (0.25) | 1.57*** (0.21) |
| Sample FE | True | True | True | True | True |
| Observations | 762 | 762 | 762 | 762 | 762 |
| Adjusted $R^2$ | 0.05 | 0.11 | 0.18 | 0.18 | 0.31 |
| F statistic | 6.63*** | 14.04*** | 23.77*** | 24.59*** | 41.91*** |

*Note:* *** $p < 0.001$; ** $p < 0.01$; * $p < 0.05$.

## 6. Discussion

### 6.1. Implications of generally low levels of student privacy concerns

The descriptive results show that students are generally not too concerned about their privacy in learning analytics (the average response on the subscale is neutral in most countries), and specifically about a possible loss of their privacy resulting from information disclosure to the university. This finding echoes Slade et al. (2019) who also found that students showed relatively low levels of concern about the pervasive collection and use of their personal data. This suggests that, based on the results of the present study, there is a receptive environment for implementations of learning analytics services in higher education across countries. While we did not collect information about how aware students are of how their data are collected, stored, and analyzed, Korir et al. (2022) found that students were more comfortable with the collection, use, and sharing of their data in the university context than in e-commerce settings. We see this reflected in both moderately high levels of trusting beliefs and low inclination towards non-self-disclosure behaviors, which echoes findings from Slade et al. (2019) that students at the Open University in the UK exhibited an inherent trust in their university to use their data appropriately and ethically. Similarly, Jones and Afnan (2019) found that US university students have a high degree of trust in their institution.

We did not examine other factors that may influence learning analytics acceptance in this study, such as students' perceived value of learning analytics. Nonetheless, it is critical to make such information available, transparent and comprehensible to students, and overall, to recentre students "as data owners rather than data objects" (Broughan & Prinsloo, 2020, p.617). This aligns with recent efforts in human-centered learning analytics (e.g., Buckingham Shum et al., 2019). How to provide transparency and clear explanations about learning analytics to students in a comprehensible way is an active and growing area of research (e.g., Jivet al., 2020; Tsai et al., 2020) that requires further investigations in different educational contexts. While students appear to generally trust their institutions, trust that grows over time can be lost quickly (Head & Hassanein, 2002; Robinson, 2020). While our study indicates overall high levels of trusting beliefs, maintaining and validating students' trusting beliefs over time is an important consideration for learning analytics designers and policy makers.

## 6.2 Implications of sample-based variation in student privacy concerns

Results of this study show that students' privacy concerns, including its two antecedents (perceived privacy control and perceived privacy risk) and two behavioral outcomes (trusting beliefs and non-self-disclosure behavior) significantly differ across some samples but not others. Students in German and Swedish samples are found to have similar levels of privacy concern, trusting beliefs, and non-self-disclosure behaviors. Likewise, Korean, Spanish and US students in our sample report similar attitudes for these constructs. Privacy concerns are lower and trusting beliefs higher among German and Swedish students than Korean, Spanish, and US students. As previously argued by Hoel and Chen (2018), students may conceptualize privacy differently across countries. In the Swedish context, trust, transparency, and openness are shared cultural values that unite Nordic nations (Nordic Council of Ministers, 2020; Smith et al., 2003). As stressed by Robinson (2020), trust in human relationships and humans' relation to technology "is a deeply ingrained cultural trait in Nordic [...] societies, where studies have shown that the *majority* of citizens trust others [...], to the extent that trust might be taken for granted in these countries" (p.1). Germany is geographically and culturally proximate to Nordic societies, which may explain the similarity in student responses. However, the basis for the similarity of responses from students in Spain, Korea, and the US for privacy concern and trusting beliefs is not clear without further consideration of cultural tendencies in the next subsection.

For the designers of learning analytics tools, it is critical to consider how students' privacy concerns can be addressed and their trust can be increased in different countries and sustained to a degree that facilitates successful implementation and adoption of targeted learning analytics services. Strategies to buttress trust in learning analytics systems may differ across countries depending on whether trust is a highly anticipated societal value. In countries where the level of students' trusting beliefs is lower (e.g., the US in our sample), designers need to consider how to raise trusting beliefs by, for example, addressing the perceived higher privacy risks. This can be achieved by several means, including the employment of privacy-by-design and privacy-by-default techniques (Cavoukian, 2006), as well as through the application of culture-sensitive (Van Boejen & Zijlistra, 2020), value-based design approaches (Chen & Zhu, 2019; Friedman & Hendry, 2019), and through trust by co-designing mechanisms (see e.g., Ahn et al., 2021) as well as the application of an interdisciplinary approach using the example of learning diaries, suggested by Veljanova et al. (2022). Furthermore, learning analytics policy makers should carefully and explicitly reflect upon values such as trust and privacy when authoring learning analytics policies, since these values may have different meanings and legal standings across countries (Egetenmeir & Homme, 2020; Hoel & Chen, 2016).

## 6.3. Implications of cultural differences in student privacy concerns

We found evidence at the individual level that culture affects all the constructs in the SPICE model of students' privacy concerns in learning analytics (Mutimukwe et al., 2022), echoing prior work that found cultural differences in privacy concerns in information systems (Milberg et al., 1995, 2000). Specifically, individual-level *power distance, masculinity* and *uncertainty avoidance* were significant predictors of students' privacy concerns. However, we did not find that *individualism (vs. collectivism)* influences privacy concerns in learning analytics, even though this cultural dimension was found to be associated with individuals' privacy concerns in other related contexts such as information systems (Cho et al., 2009; Milberg et al., 2000). This discrepancy may be explained by a difference in measurement instruments for both cultural values and privacy concerns. Students' cultural values were measured at the individual level instead of the national index reported by Hofstede et al. (2010), yielding somewhat different relative scores. For example, we found low power distance values for the US and South Korea (Table 3), whereas Hofstede et al. (2010) reported high power distance scores. Besides the difference in measurement, the discrepancy may be explained by the fact that we studied individuals' privacy concerns and cultural values in a new context, the setting of learning analytics in higher education, which has not been examined thus far.

*Power distance* is a significant positive predictor of students' perceived privacy control and perceived privacy risk (Table 4). In societies with lower *power distance* (i.e., teachers treat students as equals), we would expect students to have more perceived control over how their data is collected, stored and analysed. *Power distance* was also found to predict students' non-self-disclosure behavior in learning analytics. This suggests that students in countries with high power distance would be more hesitant to disclose their personal data to the university or other external agents, such as the LMS provider. The country-level analysis of culture also visually indicates that power distance might be the most influential cultural dimension for student privacy concerns and its outcomes, based on the slopes of the lines fitted in Figure 3. In the context of data privacy between students and institutions, it makes sense for power distance, which accounts for people's beliefs about how power should be distributed, to emerge as an important cultural difference.

*Uncertainty avoidance* is another significant predictor of students' non-self-disclosure behaviors in learning analytics, meaning that a stronger desire for uncertainty avoidance is associated with a stronger inclination towards self-disclosure. This finding supports earlier empirical research in both information systems and management (Liu & Wang, 2018; Milberg et al., 2000) and Communication Privacy Management theory that argues that individuals make self-disclosure decisions based on several criteria, including culture (Petronio, 2002). In the context of this study, we found that students reported generally high levels of *uncertainty avoidance* across all five countries, and students in the South Korean sample scored the highest (Table 3). Students with high *uncertainty avoidance* scores may expect learning analytics projects to have clear written instructions, rules and privacy regulations and governance to minimize the unpredictability in their lives and to prevent ambiguous situations (Lim et al., 2004; Smith, 1992; Vatrapu, 2011).

Higher *uncertainty avoidance* was also found to be a significant positive predictor of students' trusting beliefs in the institution for the collection and use of their data. This aligns with previous research focusing on the examination of selected cultural values on individuals' trusting beliefs in other settings (e.g., Doney et al., 1998; Hwang & Lee, 2012). Schumann et al. (2010) stress that the effect of predictability of trust should be especially high in high uncertainty avoidance cultures, which are characterized by the need for predictability, strict rules and regulations (Hofstede, 2001). The predictability of trust is thus an important consideration for learning analytics researchers and practitioners.

The cultural dimension of *masculinity* was found to positively predict students' non-self-disclosure behavior but negatively their trusting beliefs. Students representing values of masculine cultures are therefore more likely to hesitate disclosing their personal information to the university or some other external agent. This observation echoes Milberg et al.'s (2000) findings, though they did not explain this result, they conclude that masculine cultures "place greater emphasis on achievement and material success, and perhaps the economic benefits of using private information, over caring relationships and quality of life" (p.315). In response to this, Lowry et al. (2011) suggested that "to achieve work goals, highly masculine individuals may understand the need to forego a certain amount of privacy; conversely, highly feminine individuals are less achievement oriented, less competitive, and may have greater information privacy concerns" (p.174). The findings of the present study indicate overall low levels of masculinity across all five countries and low levels of both non-self-disclosure behaviors and perceived privacy control. This suggests that students who share the values of feminine cultures, as is the case in our samples, are willing to disclose their personal data to the university or other external agent. This contradicts the above-mentioned explanation by Lowry et al. (2011). Yet, even though they are willing to share their personal information, they still perceive a limited level of privacy control. This is important for learning analytics practitioners and researchers to consider to increase students' perceived privacy control. There are few examples of studies that involve stakeholders, including students, as early as possible in the design process of learning analytics systems to protect their privacy and enhance their agency (Ahn et al., 2021; Ifenthaler & Schumacher, 2016).

### 6.4 Limitations

There are several limitations to consider when drawing implications from the results of this study. First, cross-cultural survey research like the current study compares samples that differ not only in terms of their geographic location and culture, but also in terms of other, often unobservable characteristics. While we intentionally sampled similar student populations, this cannot account for a variety of local influences unrelated to the broader cultural differences. For example, our samples were drawn from institutions with a largely technical profile (i.e., STEM disciplines). We also used a non-probability sampling approach to recruit students due to the complexity and cost associated with obtaining probability samples of students across five countries. This limits the external validity of our results when it comes to understanding the average student in each country. Thus, further studies are needed to argue external validity of our findings to the broader population of students in each country. The analysis of culture at the individual level addresses this concern to an extent by characterizing culture with a fine-grained approach. Second, while most survey measures of culture exhibited sufficient internal reliability in each sample, some were too unreliable to yield robust results (in particular, the masculinity scale in South Korea). Third, while the translation of the survey instrument from English into Korean, Spanish, and Swedish was performed by native speakers and doubled or triple checked by domain experts, the translated surveys did not undergo formal evaluation. This type of limitations in cross-cultural research has been also stressed by He and van de Vijver (2011). Fourth, the responses from our study participants in each country most likely do not accurately represent that country's cultural values. Therefore, in this study we focus on their individual-level cultural values to examine cross-cultural differences. Finally, while we sampled similar student populations from selected higher educational institutions in most countries, participants recruited in Germany were approached differently and included students enrolled at many institutions, which may have affected the results.

### 6.5 Future research

It is important to note that our work does not fully bridge the gap between students' privacy concerns and cultural values, which warrants further examination with empirical research and in the design of learning analytics systems. In this study, students' privacy concerns have been studied through the lens of the SPICE model (Mutimukwe et al., 2022). However, learning analytics researchers may consider other theoretical models and constructs, including both antecedents and outcomes that may further deepen our understanding of the complex nature of students' privacy concerns across countries (e.g., Belanger & Grossler, 2011; Dinev & Hart, 2004; Smith et al., 2011). To better understand the links between individual students' cultural values and their privacy concerns in learning analytics, we recommend a qualitative research approach to complement our initial results. Finally, to understand the contextual nature of privacy in learning analytics, we call for similar studies based on data from different countries, ideally with nationally representative samples, regulatory systems, and types of higher education institutions.

### 7. Conclusion

This study shows that culture, and specifically the dimensions of *power distance*, *uncertainty avoidance*, and *masculinity* measured at the individual level affect students' privacy concerns about learning analytics in higher education across countries. The findings demonstrate that students' cultural values, especially power distance, uncertainty avoidance, and masculinity affect specific constructs in the SPICE model. We do not find any evidence that individualism is an important factor in student privacy concerns. The paper contributes to the literature by directly relating students' privacy concerns in learning analytics to cultural factors that have hitherto been underexplored by the learning analytics community, but that are generally considered influential for understanding and addressing individuals' privacy concerns in information systems. This is an important contribution to learning analytics research and practice as it highlights the need to carefully consider students' cultural values when attempting to address their privacy concerns about learning analytics.

**Appendix. Survey Instrument.**

*Introduction*

Dear participant,
Thank you for taking the time to complete our survey! You are invited to participate in this survey that aims at understanding the privacy concerns of students and their cultural values that may influence such concerns in relation to Learning Analytics (LA). LA refers to the measurement, collection, analysis and reporting of (digital) data about learners and their contexts, for purposes of understanding and optimizing learning and the environments in which it occurs.

Although collecting, analysing and reporting learning data back to the learners and teachers can improve learning and teaching, there are also potential risks to students' privacy when it comes to the handling of their personal information. One such risk is that companies or institutions might use the students' personal information for other purposes than the ones originally communicated to the students when collecting the data, violating students' right to privacy. Therefore, access and use of student information raise privacy concerns that need to be addressed if learning analytics systems are to be implemented in higher education.

*Personal information* refers to any information that can be used to identify an individual. Examples of personal information include the name, the home address, the identification card number, the location data, the IP address, etc. In the case of (online) learning, students would not disclose directly identifiable information. But, through interaction with the instructors and Learning Management Systems (e.g., Canvas or Moodle) they can provide pieces of information that can lead others to identifiable information.

*Information privacy concerns* refer to individuals' concerns about the possible loss of privacy as a result of information disclosure to an external agent/institution. They reflect an individual's perception of their concerns and worries for how their personal information is handled by a specific institution.

Earlier research has shown that cultural values may influence how individuals perceive information privacy in different contexts. Therefore, in this study, we also aim to understand whether students' cultural values may affect their information privacy concerns in relation to learning analytics and how universities handle students' personal information.

**Who are we?**

We are an international group of researchers interested in privacy issues and cultural differences in the adoption of learning analytics around the world. This research is carried out in cooperation with KTH Royal Institute of Technology (Sweden), German Institute for Educational Research (DIPF), the Ruhr-University Bochum as well as the Goethe University Frankfurt (Germany), Universidad Politécnica de Madrid (Spain), Cornell University (the US), and the Korea Advanced Institute of Science and Technology (KAIST).

**Your rights as a participant**

Participation in this questionnaire is voluntary. You have the right to cancel your participation at any time and to delete your contributions without any consequences in any form. You have the possibility to obtain information about the personal data stored by us at any time. You can at any time request that this data be corrected and deleted. You have the right at any time to demand a restriction on the processing of your data, to object to its further processing or to assert your right to data transferability. If you refuse to participate or revoke or restrict your consent, you will not face any consequences.

**How is your data collected?**

We are using different platforms in different countries to collect answers to this survey. For the subsequent evaluation of the data, all data that could lead to an identification of your person will be removed after the data collection has been completed. All collected data will be deleted after completion of the study, but at the latest after a five-year retention period. The data collected will be accessible only to researchers associated with this research project. Your answers will be evaluated anonymously and confidentially by the research team. We do not anticipate any risks from participating in this research.

The survey consists of a few demographic questions and a series of short questions and statements that you are asked to rate according to the **5-Likert scale:**

1 - strongly disagree, 2 - disagree, 3- neither agree nor disagree, 4- agree, 5 strongly agree

**Your consent**

I am asked to participate in a questionnaire about my information privacy concerns and cultural values in learning analytics. I had the opportunity to ask questions. All my questions have been answered to my satisfaction. By continuing, I confirm that I have read and understood the information above. By selecting "yes", I give my consent and am willing to participate in this research study, free from coercion and undue influence, subject to my specific approvals as stated above.

**Part 1**
1. I have understood the information provided and am participating in the study voluntarily. My information may be used for scientific purposes
    1=agree
    2=disagree
2. Gender
    1=female
    2=male
    3=other
    4=do not want to answer
3. Age (in years)
    1= 18-24 years old

        2= 25-34 years old
        3= 35-44 years old
        4= 45-54 years old
        5= 55+ years old
4. What type of study program are you enrolled in?
        1= Independent Course
        2= Bachelor's degree program
        3= Master's degree program
        4= other, please specify
5. What is your field of study?
- Humanities, Arts, Social Sciences: Anthropology, Archaeology, Economics, Education, Geography, History, Law, Linguistics, Politics, Psychology and Sociology.
- STEM: Natural Science, Technology, Engineering, Mathematics
- Medical Studies: Medical, Pharmacy, Dentistry
- Other (input)

6. Current country of residence (input)
7. Nationality (input)

**Part 2**

*Perceived privacy control*
8. I believe I have control over who can get access to my personal information, collected by my university.
9. I think I have control over *what* personal information is shared with others (e.g., third parties) by my university.
10. I believe I have control over *how* my personal information is used by my university.
11. I believe I can control my personal information that I provide to my university.

*Privacy concerns*
12. I am concerned that the information I provide to my university could be misused.
13. I am concerned that others (e.g., a person or entity who is different from oneself) can find private information about me in learning management systems.
14. I am concerned about providing personal information into learning management systems because of what others might do with it.
15. I am concerned about providing personal information when interacting with learning management systems, because it could be used in a way I did not foresee.

*Perceived privacy risk*
16. In general, it would be risky to give personal information to the learning management system used by my university.
17. There would be high potential for privacy loss associated with giving personal information to the learning management system used by my university.
18. Personal information could be inappropriately used by my university.
19. Providing my personal information to the learning management system used by my university would involve many unexpected problems.

*Trusting beliefs*
20. I trust that my university tells the truth and fulfils promises related to my personal information.
21. I trust that my university keeps my best interests in mind when dealing with personal information.
22. I trust that my university is in general predictable and consistent regarding the usage of student personal information.
23. I trust that my university is always honest with me (student) when it comes to using the information that I would provide.

*Non-self-disclosure behavior*
24. I refuse to use the learning management system because I disagree with the way my university uses personal information.
25. I take action to have my name removed from the direct mail list for the learning management system used by my university.
26. I do not want to use the learning management system employed by my university because I do not want to provide certain kind of personal information.
27. I refuse to give personal information to my university.

## Part 3

*Power distance*
27. People in higher positions should make most decisions without consulting people in lower positions.
28. People in higher positions should not ask the opinions of people in lower positions too frequently.
29. People in higher positions should avoid social interactions with people in lower positions.
30. People in lower positions should not disagree with decisions by people in higher positions.
31. People in higher positions should not delegate important tasks to people in lower positions.

*Individualism (vs collectivism)*
32. Individuals should sacrifice self-interest for the group.
33. Individuals should stick with the group even through difficulties.
34. Group welfare is more important than individuals' rewards.
35. Group success is more important than individual success.
36. Individuals should only pursue their goals after considering the welfare of the group.
37. Group loyalty should be encouraged even if individual goals suffer.

*Masculinity (vs femininity)*
38. It is more important for men to have a professional career than it is for women.
39. Men usually solve problems with logical analysis; women usually solve problems with intuition.
40. Solving difficult problems usually requires an active, forcible approach, which is typical for men.
41. There are some jobs that a man can always do better than a woman.

*Uncertainty avoidance*
42. It is important to have instructions spelled out in detail so that I always know what I am expected to do.
43. It is important to closely follow instructions and procedures.
44. Rules and regulations are important because they inform me of what is expected of me.
45. Standardized work procedures are helpful.
46. Instructions for operations are important.

47. If you have anything to add, please do it here.